\PassOptionsToPackage{dvipsnames}{xcolor}

\documentclass[letterpaper, 10 pt, conference]{ieeeconf}  % Comment this line out if you need a4paper

\IEEEoverridecommandlockouts                              % This command is only needed if 
\usepackage{amsmath,amssymb}
\usepackage{mathtools}
\usepackage{xcolor}
\usepackage{amssymb}
\usepackage{amsfonts}
\usepackage{booktabs}
\usepackage{units}
\usepackage{xcolor}
\usepackage{comment}
\usepackage{caption}
\usepackage{hyperref}
\hypersetup{
    colorlinks=true,
    linkcolor=blue,     
    urlcolor=blue,
}

\usepackage{tikz}
\usepackage{pgfplots}
\usepgfplotslibrary{groupplots} % for grouping plots
\usetikzlibrary{calc} % for calculations within TikZ
\pgfplotsset{compat=1.18}
\usetikzlibrary{decorations.pathreplacing,decorations.markings}
\usetikzlibrary{positioning}
\usetikzlibrary{backgrounds,scopes}
\usetikzlibrary {arrows.meta}
\usepackage{./results/myplotconfig} % Include your configuration
\definecolor{myRedA}{RGB}{204, 31, 31}
\definecolor{myRedB}{RGB}{204, 80, 31}
\definecolor{myRedC}{RGB}{212, 87, 87}
\definecolor{myRedD}{RGB}{135, 12, 47}

\definecolor{myOrangeA}{RGB}{227, 125, 16}
\definecolor{myOrangeB}{RGB}{219, 155, 86}
\definecolor{myOrangeC}{RGB}{219, 179, 86}
\definecolor{myOrangeD}{RGB}{242, 201, 107}

\definecolor{myGreenA}{RGB}{97, 179, 43}
\definecolor{myGreenB}{RGB}{43, 179, 63}
\definecolor{myGreenC}{RGB}{119, 237, 136}
\definecolor{myGreenD}{RGB}{155, 219, 7}

\definecolor{myBlueA}{RGB}{7, 219, 173}
\definecolor{myBlueB}{RGB}{7, 208, 219}
\definecolor{myBlueC}{RGB}{7, 138, 219}
\definecolor{myBlueC}{RGB}{7, 67, 219}

\definecolor{myPurpleA}{RGB}{116, 80, 173}
\definecolor{myPurpleB}{RGB}{62, 4, 128}
\definecolor{myPurpleC}{RGB}{118, 4, 224}
\definecolor{myPurpleC}{RGB}{204, 127, 245}

\definecolor{myPinkA}{RGB}{220, 131, 230}
\definecolor{myPinkB}{RGB}{214, 6, 207}

\definecolor{myGray}{RGB}{118, 110, 108}

\newcommand{\R}{\mathbb{R}}
\newcommand{\N}{\mathbb{N}}

\newcommand{\mbb}[1]{\mathbb{#1}}
\newcommand{\mrm}[1]{\mathrm{#1}}
\newcommand{\mc}[1]{\mathcal{#1}}
\newcommand{\diag}[1]{\mrm{diag}\big({#1}\big)}

\newtheorem{defi}{Definition}[section]

\newtheorem{prop}[defi]{Proposition}
\newtheorem{lemma}[defi]{Lemma}
\newtheorem{problem}[defi]{Problem}

\newtheorem{assumption}[defi]{Assumption}
\newtheorem{condition}[defi]{Condition}

\pgfplotsset{compat=1.18}

\let\oldbibliography\thebibliography
\renewcommand{\thebibliography}[1]{%
  \oldbibliography{#1}%
  \setlength{\itemsep}{-.7pt}%
}
 
\title{\LARGE \bf
Model Predictive Control of District Heating Grids Using Stabilizing Terminal Ingredients
}

\author{Max Rose$^{1}$, Hannes Gernandt$^{1}$, Juan E. Machado$^{2}$, Johannes Schiffer$^{1,2}$ % <-this % stops a space
\thanks{$^{1}$Fraunhofer Research Institution for Energy Infrastructures and Geothermal Systems IEG, 03046 Cottbus, Germany, {\tt\small \{max.rose, hannes.gernandt, johannes.schiffer\}@ieg.fraunhofer.de}}%
\thanks{$^{2}$Brandenburg University of Technology Cottbus-Senftenberg BTU, 03046 Cottbus, Germany, {\tt\small \{machadom, schiffer\}@b-tu.de}}%
}

\begin{document}
\topskip=2mm
\floatsep=3mm
\textfloatsep=3mm
\columnsep=2mm
\abovedisplayskip=1mm
\belowdisplayskip=1mm
\arraycolsep=.75mm

% make the title area
\maketitle
\thispagestyle{empty}
\pagestyle{empty}

% As a general rule, do not put math, special symbols or citations
% in the abstract
\begin{abstract}
The transformation of fossil fuel-based district heating grids (DHGs) to CO$_2$-neutral DHGs requires the development of novel operating strategies.
Model predictive control (MPC) is a promising approach, as knowledge about future heat demand and heat supply can be incorporated into the control, operating constraints can be ensured and the stability of the closed-loop system can be guaranteed.
In this paper, we employ MPC for DHGs to control the system mass flows and injected heat flows.
Following common practice, we derive terminal ingredients %, i.e., additional costs and constraints that are added to the nominal optimization problem of the MPC, 
to stabilize given steady state temperatures and storage masses in the DHG. 
%To this end, we consider DHGs consisting of a network of pipelines that transport water as energy carrier in between producers, consumers and thermal energy storages.
%By introducing fundamental mass flows, the DHN's flow dynamics can be rewritten as a set of ordinary differential equations. 
To apply MPC with terminal ingredients, it is crucial that the system under control is stabilizable.
By exploiting the particular system structure, we give a sufficient condition for the stabilizability in terms of the grid topology and hence, for the applicability of the MPC scheme to DHGs. 
Furthermore, we demonstrate the practicability of the application of MPC to an exemplary DHG in a numerical case study.
\end{abstract}

% no keywords

\section{Introduction}
%Mitigating the effects of anthropogenic global climate change is a shared political objective in numerous countries, closely aligned with the pursuit of CO$_2$-neutrality in the energy sector, primarily through the utilization of renewable energy sources (RESs) \cite{NetZero2021}. 
%Notably, within the energy sector of 28 European countries, 50\% of final energy consumption is used for heating purposes \cite{trier_guidelines_2018}.
%Thus, decarbonizing the heat sector is a major target to reduce global climate change.
%
%In this context, district heating grids (DHGs) play an important role \cite{LUND2014131}. 
District heating grids (DHGs) play an important role in the decarbonization of the heat sector \cite{LUND2014131}.
A DHG is a network of pipes, which are used to transport water as heat energy carrier in between producers, where heat flows are injected, and consumers, where heat flows are extracted with individual temperature requirements. 
Major advantages of DHGs are the possibility of a decentralized integration of RES-based heat producers, low temperature waste heat sources and thermal energy storages (TESs), which are all needed for the transformation towards RES-based DHGs \cite{trier_guidelines_2018,LUND2014131}.
This transformation influences the operation of future DHGs, such that conventional operating strategies are not feasible for an RES-based operation of DHGs \cite{LUND20141}.
This motivates the development of novel operating strategies for RES-based DHGs.

To give a brief overview of the state of the art in the field of operational management of DHGs, we distinguish between energy management strategies (EMSs), which compute control signals by solving an optimization problem iteratively, and control approaches, which ensure the stability of the closed-loop system by a suitable feedback controller.

An example of economically motivated EMS for DHGs can be found e.g. in \cite{rose_2023}, where the flexibility stemming from sector coupling is used to increase operational efficiency.
In \cite{krug_nonlinear_2021}, suitable discretization schemes to obtain a nonlinear optimization model for DHGs with increased modeling accuracy are investigated.
A reduced order model of a DHG is investigated to reduce computational complexity in~\cite{rein_optimal_2020}.

Control strategies for the nonlinear hydraulics of DHGs are studied e.g. in \cite{machado_adaptive_2022}, where a passivity-based approach is proposed.
A~temperature controller based on Lyapunov-Krasovskii stability theory is proposed and experimentally validated in \cite{bendtsen_control_2017}.
In~\cite{scholten_modeling_2015}, the mass stored in a TES as well as the system temperatures are controlled in a DHG with a single producer and multiple consumers.
Furthermore, passivity-based approaches are proposed in \cite{machado_modeling_2022} and \cite{Krishna.2021} to control the thermo-hydraulics of DHGs and to control the frequency and temperature in a coupled power and heating grid, respectively.

To combine the advantages from both optimization-based EMS, e.g. a predictive operating behavior as well as constraint satisfaction, and control strategies guaranteeing closed-loop stability, model predictive control (MPC) is a highly promising approach.
In MPC, an optimization problem is solved iteratively to obtain a stabilizing control input for each control action \cite{GrunePannek17}.
However, none of the above papers addresses an optimization-based controller with assured stability for DHGs.

In \cite{chen_quasi-infinite_1998}, a set constraint, so called terminal constraint, and a cost term, so called terminal cost, was added to the nominal optimization problem of a MPC to obtain an asymptotically stable closed-loop behaviour.
The terminal constraint is defined via a neighborhood around the desired steady-state set point, called terminal region, and enforces the last entry of each optimal state trajectory resulting from solving the optimization problem to lie in this region.
Additionally, the terminal cost penalizes the distance of the last entry of each optimal state trajectory to the desired steady-state set point. 
%Nevertheless, a local controller based on the linearization of the system at the desired steady state is required to obtain the stabilizing terminal cost and terminal region.
Together, terminal region and terminal cost form the terminal ingredients.
The approach from \cite{chen_quasi-infinite_1998} is extended to the discrete-time case in \cite{Rajhans2017}.
Additionally, \cite{chen_quasi-infinite_1998,Rajhans2017} provide a framework to derive a terminal cost and a terminal region.
However, to apply MPC with terminal ingredients, it is crucial that the linearized system model is stabilizable with respect to the desired steady state.
Furthermore, the algorithm to calculate terminal cost and terminal constraint can result in conservative MPC laws \cite{chen_quasi-infinite_1998}.
This motivates both the examination of the local stabilizability of a DHG model as well as a numerical investigation of the practicability of MPC with terminal ingredients conducted in the present work.

Driven by the importance of RES-based DHG operation in the transition to a zero-emission heating sector and recognizing the pivotal role of MPC in this context, this paper presents three primary contributions:
At first, building upon \cite{krug_nonlinear_2021,machado_modeling_2022}, 
we develop, for a DHG with multiple producers, multiple distributed storage units and consumers, an ordinary differential equation-based (ODE-based) state model that describes relevant mass and temperature dynamics and which can be used to implement an MPC.
Compared to \cite{krug_nonlinear_2021,machado_modeling_2022}, we introduce heat losses from \cite{krug_nonlinear_2021} into the modeling approach in \cite{machado_modeling_2022} and allow for a less restrictive placement of TES which were not considered in  \cite{krug_nonlinear_2021}. 
Additionally, multiple temperature layers are considere in this work.
Second, we derive a sufficient condition for the local stabilizability of the developed DHG model by exploiting its structural properties.
Third, we calculate suitable terminal ingredients for an exemplary DHG and demonstrate the practicability of MPC with terminal ingredients to control DHGs in a case study.

The remainder of the paper is structured as follows.
In Section \ref{sec: model}, we describe the DHG model.
Then, in Section \ref{sec: MPC}, we formulate the optimization problem of the MPC, show the stabilizability property of the system and derive a stabilizing control law.
A case study demonstrating the practicability of the approach is presented in Section \ref{sec: case study}.
The presented work is summarized and conclusions are drawn in Section \ref{sec: conclusion}.

\subsection*{Notation}
The set of (positive and negative) real numbers and non-negative integers is denoted by ($\R_+$ and $\R_-$) $\R$ and $\N$, respectively.
%Additionally, $\R_+$ and $\R_-$ denote the set of positive and negative real valued numbers, respectively.
For a finite set $\mc{S}$, $|\mc{S}|$ denotes its cardinality.
For a vector $x \in \R^n$, $\diag{x} \in \R^{n \times n}$ denotes a diagonal matrix having the elements of $x$ on the main diagonal.
The weighted norm $x^\top Q x$, where $Q \in \R^{n \times n}$, is denoted by $\|x\|_Q^2$.
Let $X \in \R^{n \times m}$ denote a $n \times m$ matrix and let $\mc{P}$ denote a set with $|\mc{P}|=n$, in which each element of $\mc{P}=\{p_1,\ldots,p_n\}$ is associated with one row of $x$ or $X$ and $\mc{P}_{\mrm{sub}} \subset \mc{P}$. 
Then, $(x)_{p \in \mc{P}_{\mrm{sub}}} \in \R^{|\mc{P}_{\mrm{sub}}|}$ and $(X)_{p \in \mc{P}_{\mrm{sub}}} \in \R^{|\mc{P}_{\mrm{sub}}| \times m}$ denote a vector and a matrix comprising only those rows of $x$ and $X$, respectively, for which $p \in \mc{P}_{\mrm{sub}}$, and $(x)_{p_i} \in \R$ denotes the entry of $x$ that is associated with $p_i \in \mc{P}$.
The $n \times n$ identity matrix is denoted by $I_n$.
Further, $|X| \in \R^{n \times m}$ is the matrix resulting from $X$ by taking absolute values of its entries.

\section{Thermo-hydraulic model} \label{sec: model}
\begin{figure}[!htbp]
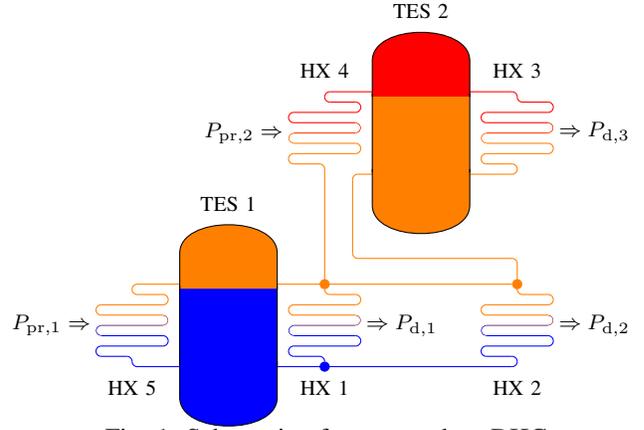

    \centering
    \usetikzlibrary{decorations.pathreplacing,decorations.markings}
\usetikzlibrary{positioning}
\usetikzlibrary{backgrounds,scopes}
\usetikzlibrary {arrows.meta}
\usetikzlibrary{patterns}
\usetikzlibrary{fadings}

\newcommand{\unitSize}{10mm}
\newcommand{\myScaling}{.4}
\newcommand{\myFactor}{\unitSize*\myScaling}

\newcommand{\dx}{1.2}
\newcommand{\dy}{.75}
\newcommand{\dr}{1.2/7}

\newcommand{\dX}{3.2}
\newcommand{\dY}{4}

\pgfdeclarelayer{bg}    % declare background layer
\pgfsetlayers{bg,main}  % set the order of the layers (main is the standard layer)

\input{figures/networkUnits}
\input{results/myColors}

\begin{tikzpicture}
    % Add producing HXs
	\node[label={[label distance=-.45*\unitSize*\myScaling]180:\footnotesize$P_{\mathrm{pr,}1} \Rightarrow$}, label={[label distance=-.0*\unitSize*\myScaling]270:\footnotesize HX 5}] (HXprA)  at  (0, 0)     {\HXlayerA{\unitSize}{\myScaling}}; 
    \node[label={[label distance=-.45*\unitSize*\myScaling]180:\footnotesize$P_{\mathrm{pr,}2} \Rightarrow$}, label={[label distance=-.0*\unitSize*\myScaling]90:\footnotesize HX 4}] (HXprB)  at  (2*\dX*\myFactor, 1.6*\dY*\myFactor)     {\HXlayerB{\unitSize}{\myScaling}};
    
    % Add consuming HXs
    \node[label={[label distance=-.45*\unitSize*\myScaling]0:\footnotesize$\Rightarrow P_{\mathrm{d,}1}$}, label={[label distance=-.0*\unitSize*\myScaling]270:\footnotesize HX 1}] (HXdA)  at  (2*\dX*\myFactor, 0)     {\HXlayerA{\unitSize}{\myScaling}};
    \node[label={[label distance=-.45*\unitSize*\myScaling]0:\footnotesize$\Rightarrow P_{\mathrm{d,}2}$}, label={[label distance=-.0*\unitSize*\myScaling]270:\footnotesize HX 2}] (HXdB)  at  (4*\dX*\myFactor, 0)     {\HXlayerA{\unitSize}{\myScaling}};
    \node[label={[label distance=-.45*\unitSize*\myScaling]0:\footnotesize$\Rightarrow P_{\mathrm{d,}3}$}, label={[label distance=-.0*\unitSize*\myScaling]90:\footnotesize HX 3}] (HXdC)  at  (4*\dX*\myFactor, 1.6*\dY*\myFactor)     {\HXlayerB{\unitSize}{\myScaling}};
    
    % Add TESs
    \node[label={[label distance=-0.5*\unitSize*\myScaling]90:\footnotesize TES 1}] (TESA)    at  (\dX*\myFactor, 0)   {\TESlayerA{\unitSize}{\myScaling}};  
    \node[label={[label distance=-0.5*\unitSize*\myScaling]90:\footnotesize TES 2}] (TESB)    at  (3*\dX*\myFactor, 1.6*\dY*\myFactor)   {\TESlayerB{\unitSize}{\myScaling}};

    % Add pipe intersections
    \filldraw[color = orange, fill = orange] (2*\dX*\myFactor, 0+\dr*\myFactor+\dx*\myFactor) circle (.15*\myFactor);
    \filldraw[color = blue, fill = blue] (2*\dX*\myFactor, 0-\dr*\myFactor-\dx*\myFactor) circle (.15*\myFactor);
    \filldraw[color = orange, fill = orange] (4*\dX*\myFactor, 0+\dr*\myFactor+\dx*\myFactor) circle (.15*\myFactor);
    
    \begin{pgfonlayer}{bg}
        % Connect HXprA to TES1
        \draw[orange] (0, 0+\dx*\myFactor) to[out=90, in=180] (0+\dr*\myFactor, 0+\dr*\myFactor+\dx*\myFactor);
        \draw[orange] (0+\dr*\myFactor, 0+\dr*\myFactor+\dx*\myFactor) -- (0+\dX*\myFactor-1.35*\dx*\myFactor, 0+\dr*\myFactor+\dx*\myFactor); 
        \draw[blue] (0, 0-\dx*\myFactor) to[out=270, in=180] (0+\dr*\myFactor, 0-\dr*\myFactor-\dx*\myFactor);
        \draw[blue] (0+\dr*\myFactor, 0-\dr*\myFactor-\dx*\myFactor) -- (0+\dX*\myFactor-1.35*\dx*\myFactor, 0-\dr*\myFactor-\dx*\myFactor);

        % Connect TES1 to pipe intersection 1
        \draw[orange] (1.35*\dx*\myFactor, 0+\dr*\myFactor+\dx*\myFactor) -- (2*\dX*\myFactor, 0+\dx*\myFactor+\dr*\myFactor);
        \draw[blue] (\dX*\myFactor+1.35*\dx*\myFactor, 0-\dr*\myFactor-\dx*\myFactor) -- (2*\dX*\myFactor, 0-\dx*\myFactor-\dr*\myFactor);

        % Connect pipe intersection 2 to TES1
        \draw[orange] (2*\dX*\myFactor, 0+\dx*\myFactor+\dr*\myFactor) -- (2*\dX*\myFactor, 1.6*\dY*\myFactor-\dx*\myFactor);

        % Connect pipe intersection 2 to HXprB
        \draw[blue] (2*\dX*\myFactor, 0-\dx*\myFactor-\dr*\myFactor) -- (4*\dX*\myFactor-\dr*\myFactor, 0-\dx*\myFactor-\dr*\myFactor);
        \draw[blue] (4*\dX*\myFactor-\dr*\myFactor, 0-\dx*\myFactor-\dr*\myFactor) to[out=0, in=270] (4*\dX*\myFactor, 0-\dx*\myFactor);

        % Connect pipe intersection 1 to pipe intersection 3
        \draw[orange] (2*\dX*\myFactor, 0+\dx*\myFactor+\dr*\myFactor) -- (4*\dX*\myFactor-\dr*\myFactor, 0+\dx*\myFactor+\dr*\myFactor);

        % Connect HXprB to TES2
        \draw[red] (2*\dX*\myFactor, 1.6*\dY*\myFactor+\dx*\myFactor) to[out=90, in=180] (2*\dX*\myFactor+\dr*\myFactor, 1.6*\dY*\myFactor+\dr*\myFactor+\dx*\myFactor);
        \draw[red] (2*\dX*\myFactor+\dr*\myFactor, 1.6*\dY*\myFactor+\dr*\myFactor+\dx*\myFactor) -- (2*\dX*\myFactor+\dX*\myFactor-1.35*\dx*\myFactor, 1.6*\dY*\myFactor+\dr*\myFactor+\dx*\myFactor);

        % Connect TES2 to HXdC
        \draw[red] (3*\dX*\myFactor+1.35*\dx*\myFactor, 1.6*\dY*\myFactor+\dr*\myFactor+\dx*\myFactor) -- (4*\dX*\myFactor-\dr*\myFactor, 1.6*\dY*\myFactor+\dr*\myFactor+\dx*\myFactor);
        \draw[red] (4*\dX*\myFactor-\dr*\myFactor, 1.6*\dY*\myFactor+\dr*\myFactor+\dx*\myFactor) to[out=0, in=90] (4*\dX*\myFactor, 1.6*\dY*\myFactor+\dx*\myFactor);
        \draw[orange] (3*\dX*\myFactor+1.35*\dx*\myFactor, 1.6*\dY*\myFactor-\dr*\myFactor-\dx*\myFactor) -- (4*\dX*\myFactor-\dr*\myFactor, 1.6*\dY*\myFactor-\dr*\myFactor-\dx*\myFactor);
        \draw[orange] (4*\dX*\myFactor-\dr*\myFactor, 1.6*\dY*\myFactor-\dr*\myFactor-\dx*\myFactor) to[out=0, in=270] (4*\dX*\myFactor, 1.6*\dY*\myFactor-\dx*\myFactor);

        % Connect TES2 to pipe intersection 3
        \draw[orange] (3*\dX*\myFactor-1.35*\dx*\myFactor, 1.6*\dY*\myFactor-\dx*\myFactor-\dr*\myFactor) -- (3*\dX*\myFactor-1.75*\dx*\myFactor, 1.6*\dY*\myFactor-\dx*\myFactor-\dr*\myFactor);
        \draw[orange] (3*\dX*\myFactor-1.75*\dx*\myFactor, 1.6*\dY*\myFactor-\dx*\myFactor-\dr*\myFactor) to[out=180, in=90] (3*\dX*\myFactor-1.75*\dx*\myFactor-\dr*\myFactor, 1.6*\dY*\myFactor-\dx*\myFactor-2*\dr*\myFactor);
        \draw[orange] (3*\dX*\myFactor-1.75*\dx*\myFactor-\dr*\myFactor, 1.6*\dY*\myFactor-\dx*\myFactor-2*\dr*\myFactor) -- (3*\dX*\myFactor-1.75*\dx*\myFactor-\dr*\myFactor, 0.6*\dY*\myFactor);
        \draw[orange] (3*\dX*\myFactor-1.75*\dx*\myFactor-\dr*\myFactor, 0.6*\dY*\myFactor) to[out=270, in=180] (3*\dX*\myFactor-1.75*\dx*\myFactor, 0.6*\dY*\myFactor-\dr*\myFactor);
        \draw[orange] (3*\dX*\myFactor-1.75*\dx*\myFactor, 0.6*\dY*\myFactor-\dr*\myFactor) -- (4*\dX*\myFactor-\dr*\myFactor, 0.6*\dY*\myFactor-\dr*\myFactor);
        \draw[orange] (4*\dX*\myFactor-\dr*\myFactor, 0.6*\dY*\myFactor-\dr*\myFactor) to[out=0, in=90] (4*\dX*\myFactor, 0.6*\dY*\myFactor-2*\dr*\myFactor);
        \draw[orange] (4*\dX*\myFactor, 0.6*\dY*\myFactor-2*\dr*\myFactor) -- (4*\dX*\myFactor, 0+\dx*\myFactor);
    \end{pgfonlayer}

\end{tikzpicture}
    \vspace*{-3mm}
    \caption{Schematic of an exemplary DHG.}
    \label{fig:exmplDHG_skizze}
\end{figure}

\begin{figure}[!h]
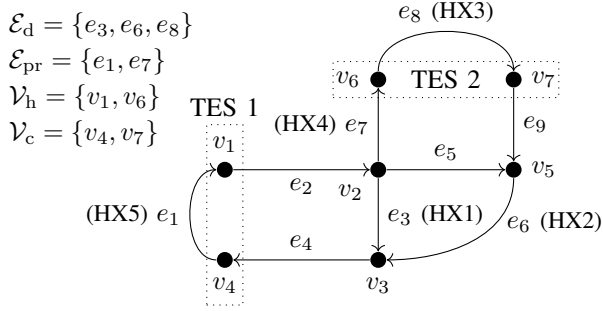

    \centering
    \usetikzlibrary{decorations.pathreplacing,decorations.markings}
\usetikzlibrary{positioning}
\usetikzlibrary{backgrounds,scopes}
\usetikzlibrary{arrows}

\newcommand{\unitSize}{10mm}
\newcommand{\myScaling}{.4}
\newcommand{\myFactor}{\unitSize*\myScaling}

\newcommand{\dX}{3}

\input{figures/networkUnits}
\input{results/myColors}

\renewcommand{\baselinestretch}{2}

\begin{tikzpicture}
	\node[circle,fill=black,inner sep=0pt,minimum size=6pt,label=above:{$v_1$}] (v1) at (-0.2*\dX*\myFactor, 0) {};
    \node[circle,fill=black,inner sep=0pt,minimum size=6pt,label=south west:{$v_2$}] (v2) at  (1.5*\dX*\myFactor, 0) {};
    \node[circle,fill=black,inner sep=0pt,minimum size=6pt,label=below:{$v_3$}] (v3) at (1.5*\dX*\myFactor, -\dX*\myFactor) {};
    \node[circle,fill=black,inner sep=0pt,minimum size=6pt,label=below:{$v_4$}] (v4) at (-0.2*\dX*\myFactor, -\dX*\myFactor) {};
    \node[circle,fill=black,inner sep=0pt,minimum size=6pt,label=right:{$v_5$}] (v5) at (3*\dX*\myFactor, 0) {};

    \node[circle,fill=black,inner sep=0pt,minimum size=6pt,label=left:{$v_6$}] (v6) at (1.5*\dX*\myFactor, \dX*\myFactor) {};
    \node[circle,fill=black,inner sep=0pt,minimum size=6pt,label=right:{$v_7$}] (v7) at (3*\dX*\myFactor, \dX*\myFactor) {};
    
    \draw[->] (v1) -- node[below] {$e_2$} (v2);
    \draw[->] (v2) -- node[right] {$e_3$ \small (HX1)} (v3);
    \draw[->] (v2) -- node[above] {$e_5$} (v5);
    \draw[->] (v2) -- node[left] {\small (HX4) \normalsize $e_7$} (v6);
    \draw[->] (v3) -- node[above] {$e_4$} (v4);
    \draw[->] (v4) to [out=180,in=180] node[left] {\small (HX5) \normalsize $e_1$} (v1);
    \draw[->] (v6) to [out=90,in=90] node[above] {$e_8$ \small (HX3)} (v7);
    \draw[->] (v7) -- node[right] {$e_9$} (v5);
    \draw[->] (v5) to [out=270,in=0] node[right, pos=0.3] {$e_6$ \small (HX2)} (v3);

    \draw[black, dotted] (-0.4*\dX*\myFactor, 0.5*\dX*\myFactor) rectangle (0.*\dX*\myFactor, -1.5*\dX*\myFactor);
    \node[align=right, label={[label distance=-4.0pt]north:{TES 1}}] at (-0.2*\dX*\myFactor, 0.5*\dX*\myFactor) {};

    \draw[black, dotted] (1.*\dX*\myFactor, 0.8*\dX*\myFactor) rectangle (3.5*\dX*\myFactor, 1.2*\dX*\myFactor);
    \node[align=center, label={[label distance=-3.5pt]:{TES 2}}] at (2.25*\dX*\myFactor, 0.8*\dX*\myFactor) {};

    \node[align=left, label=right:{$\mathcal{E}_{\mathrm{d}} = \{ e_3, e_6, e_8 \}$}] (Ed) at (-2.8*\dX*\myFactor, 1.6*\dX*\myFactor) {};
    \node[align=left, label=right:{$\mathcal{E}_{\mathrm{pr}} = \{ e_1, e_7 \}$}] (Epr) at (-2.8*\dX*\myFactor, 1.2*\dX*\myFactor) {};
    \node[align=left, label=right:{$\mathcal{V}_{\mathrm{h}} = \{ v_1, v_6 \}$}] (Vh) at (-2.8*\dX*\myFactor, 0.8*\dX*\myFactor) {};
    \node[align=left, label=right:{$\mathcal{V}_{\mathrm{c}} = \{ v_4, v_7 \}$}] (Vc) at (-2.8*\dX*\myFactor, 0.4*\dX*\myFactor) {};
    
\end{tikzpicture}
    \caption{Graph representation of the exemplary DHG from Figure \ref{fig:exmplDHG_skizze} with definitions of associated sets $\mc{E}_{\mrm{d}}$, $\mc{E}_{\mrm{pr}}$, $\mc{V}_{\mrm{h}}$ and $\mc{V}_{\mrm{c}}$. Dotted rectangles frame vertices of a corresponding TES.}
    \label{fig:exmplDHG_graph}
\end{figure}

The goal of this section is to develop an ODE-based model of a general DHG that can be used to design an MPC feedback law that stabilizes the water mass stored at TESs and the temperatures within the DHG. % by controlling the mass flows and the injected heat.
Then, the model needs to describe the hydrodynamics of water mass stored at TESs, the thermodynamics of water in the DHG components as well as Kirchhoff's laws for hydraulic networks.
To do so, we write-down the balance of mass and the balance of energy of a general DHG that consists of a network of pipes connecting stratified TESs and heat exchangers (HXs).
The HXs are associated with either an injected heat flow from a producer or an extracted heat flow from a consumer.
In Figure~\ref{fig:exmplDHG_skizze}, a schematic representation of an exemplary DHG containing two HXs associated with the injected heat flows $P_{\mrm{pr},1} \in \R_+$ and $P_{\mrm{pr},2} \in \R_+$, three HXs associated with the extracted heat flows $P_{\mrm{d},1} \in \R_+$, $P_{\mrm{d},2} \in \R_+$ and $P_{\mrm{d},3} \in \R_+$ and two TESs interconnected by a network of pipes is shown. 
\linebreak
We make the following standard assumptions to derive the DHG model \cite{Borsche19,hauschild_port-hamiltonian_2020,krug_nonlinear_2021}.
\begin{assumption} \label{ass1}
    (i)~A fixed control volume is associated with each component.
    (ii)~All components are completely filled with water at all times.
    (iii)~The water in the components is incompressible.
    (iv)~Gravitational acceleration is neglected.
    (v)~The mass flow through a pipe is one-dimensional.
    (vi)~The internal energy of water is regarded as the main source of energy, i.e., other forms of energy are negligible.
    (vii)~The internal energy of water depends linearly on the temperature of the water.
\end{assumption}
For modeling TESs, we make the following assumptions \cite{machado_modeling_2022}.
\begin{assumption} \label{ass2}
    (i)~Forces acting within a TES or between individual water layers of a TES are neglected.
    (ii)~The control volume of a TES can ideally be divided into a hot layer and a cold layer. 
\end{assumption}
Formally, we represent the topology of a DHG by a connected graph $\mc{G}=(\mc{V},\mc{E})$, where $\mc{V}=\{v_1, \ldots, v_{|\mc{V}|}\}$ and $\mc{E}=\{e_1, \ldots, e_{|\mc{E}|}\}$ denote the sets of vertices and edges, respectively.
The set $\mc{V}$ consists of three subsets, namely the set of vertices associated with pipe intersections $\mc{V}_{\mrm{p}} \subset \mc{V}$, the set of vertices associated with hot layers of a TES $\mc{V}_{\mrm{h}} \subset \mc{V}$, and the set of vertices associated with cold layers of a TES $\mc{V}_{\mrm{c}} \subset \mc{V}$, such that $\mc{V} = \mc{V}_{\mrm{p}} \cup \mc{V}_{\mrm{h}} \cup \mc{V}_{\mrm{c}}$.
An edge $e \in \mc{E}$ is associated with a pipe $\mc{E}_{\mrm{p}} \subset \mc{E}$, or a HX.
Edges associated with HXs, where heat flow is injected or heat flow is extracted, are collected in the sets $\mc{E}_{\mrm{pr}} \subset \mc{E}$ or $\mc{E}_{\mrm{d}} \subset \mc{E}$, respectively, such that $\mc{E} = \mc{E}_{\mrm{p}} \cup \mc{E}_{\mrm{pr}} \cup \mc{E}_{\mrm{d}}$.
For the following modeling, a directed graph is required, which can be obtained by assigning an arbitrary orientation to $e \in \mc{E}$.
For an intuitive physical interpretation, we choose the orientation of $e \in \mc{E}$ corresponding to the direction of flow through a pipe associated with this edge.
Then, the edge $e = (v,w) \in \mc{E}$ exists, if water flows from $v \in \mc{V}$ to $w \in \mc{V}$ for $v \neq w$.
In Figure \ref{fig:exmplDHG_graph}, the graph representation of the DHG from Figure~\ref{fig:exmplDHG_skizze} is shown.
The vertex-edge incidence matrix $B \in \R^{|\mc{V}| \times |\mc{E}|}$ of $\mc{G}$ is defined element-wise via
\begin{equation*}
    (B)_{i,j}=
	\begin{cases}
	1, & \text{if~$e_j=(w,v_i)\in\mc{E}$}, \\
	-1, & \text{if $e_j=(v_i,w)\in\mc{E}$}, \\
	0, & \text{otherwise.}
	\end{cases}
\end{equation*}

We proceed by introducing some further concepts from graph theory from \cite{bollobas_modern_1998} that are used in the following sections.
A path defines a sequence of edges that link a sequence of distinct vertices.
Let $G=(V,E)$ with set of vertices $V$ and set of edges $E=\{ \mrm{e}_1, \ldots, \mrm{e}_{|E|} \}$ denote a weakly connected graph, i.e., a graph that has an undirected path between every two vertices.
Moreover, let a cycle be an undirected sequence of consecutive vertices and edges in which only the first and last vertex are equal.
If $G$ contains cycles, removing all edges $b \in \mc{F} \subset E$ from $G$ yields a connected subgraph $S=(V,E \setminus \mc{F})$ that does not contain cycles.
Then $S$ is called the spanning tree of $G$, whereas $\mc{F}$ is called the set of chords. 
For the example given in Figure~\ref{fig:exmplDHG_graph}, $\{ e_3, e_6, e_7 \}$ would be a valid choice of chords.
A weakly connected graph has at least one spanning tree \cite{bollobas_modern_1998}.
In what follows, every set of edges $L_j \subset E$, $j=1,\ldots,|\mc{F}|$, that forms a unique cycle by adding $b \in \mc{F}$ to $S$ is called a fundamental cycle.
The orientation of a fundamental cycle is defined in accordance with the orientation of $b \in \mc{F}$ that forms it.
Now, we can define the fundamental cycle matrix $F \in \R^{|\mc{F}| \times |E|}$ of $G$ element-wise as
\begin{equation*}
    (F)_{i,j}=
	\begin{cases}
	1, & \text{if } \mrm{e}_j \in L_i \text{ is oriented as } L_i, \\
	-1, & \text{if } \mrm{e}_j \in L_i \text{ is not oriented as } L_i, \\
	0, & \text{if } \mrm{e}_j \notin L_i \text{, otherwise.}
	\end{cases}
\end{equation*}

\subsection*{Balance of mass}
We assign a mass to each component of the DHG and a~mass flow to each pipe.
Thus, vertex $v \in \mc{V}$ has a mass and each edge $e \in \mc{E}$ has a mass and a~mass flow.
The sign of a mass flow is considered to be positive in the direction of the associated edge.
From Assumption \ref{ass1} it follows that the masse of edges $e \in \mc{E}$ and vertice ${v \in \mc{V}_{\mrm{p}}}$ are constant and collected in the vectors ${(m_{\mrm{v}})_{v \in \mc{V}_{\mrm{p}}} \in \R_+^{|\mc{V}_{\mrm{p}}|}}$ and ${m_{\mrm{e}} \in \R_+^{|\mc{E}|}}$, respectively. 
%Consequently, $(m_{\mrm{e}})_{e \in \mc{E}} = ~\text{const.}$ and $(m_{\mrm{v}})_{v \in \mc{V}_{\mrm{p}}} = ~\text{const.}$, where $m_{\mrm{e}} \in \R_+^{|\mc{E}|}$ and $m_{\mrm{v}} \in \R_+^{|\mc{V}|}$ denote vectors collecting the masses from all edges and vertices, respectively.
Furthermore, from Assumption \ref{ass1} it also follows that the total mass of the $i^{\text{th}}$ TES denoted by $m_{\mrm{tes},i} \in \R_+$ is constant, while we allow for a varying ratio of hot to cold water in each TES. Thus, if the vertices $v_h \in \mc{V}_{\mrm{h}}$ and $v_c \in \mc{V}_{\mrm{c}}$ are associated with the $i^{\text{th}}$ TES, the balance of mass reads
\vspace*{-1.mm}
\begin{equation} \label{eq:mh+mc=mtes}
    m_{\mrm{tes,}i} = (m_{\mrm{v}})_{v_h}(t) + (m_{\mrm{v}})_{v_c}(t),\quad  \forall t \geq 0.
\end{equation}

We define $m_{\mrm{h}} := (m_{\mrm{v}})_{v \in \mc{V}_{\mrm{h}}}$ and $m_{\mrm{c}} := (m_{\mrm{v}})_{v \in \mc{V}_{\mrm{c}}}$ to obtain the following system of differential-algebraic equations
\begin{equation} \label{eq:DAE}
    \begin{bmatrix} 0 \\ \dot m_{\mrm{h}}(t) \\ \dot m_{\mrm{c}}(t) \end{bmatrix} = \begin{bmatrix} (B)_{v \in \mc{V}_{\mrm{p}}} \\ (B)_{v \in \mc{V}_{\mrm{h}}} \\ (B)_{v \in \mc{V}_{\mrm{c}}} \end{bmatrix} q_{\mrm{e}}(t) = 
    \begin{bmatrix} (B)_{v \in \mc{V}_{\mrm{p}}} \\ (B)_{v \in \mc{V}_{\mrm{h}}} \\ (-B)_{v \in \mc{V}_{\mrm{h}}} \end{bmatrix} q_{\mrm{e}}(t),
\end{equation}
which collect the mass balance at each vertex of the DHG, where $q_{\mrm{e}} \in \R_+^{|\mc{E}|}$ denotes a vector collecting the mass flows through all edges.
Note that $\dot m_{\mrm{h}} = -\dot m_{\mrm{c}}$ follows from \eqref{eq:DAE}.
Thus, the dynamics of the mass stored in the TESs can completely be determined by the dynamics of $\dot m_{\mrm{h}}$.
Also note, since the mass flows over the edges $q_{\mrm{e}}$ linearly depend on each other due to Kirchhoff's laws, see first line of \eqref{eq:DAE}, they can be expressed by a subset of independent variables \cite{de_persis_pressure_2011}, which we refer to as the fundamental mass flows describing the entire hydraulic state of the DHG.

For the considered DHG, we proceed to identify the fundamental mass flows following the procedure described in~\cite{machado_modeling_2022}.
In hydraulic networks whose topology can be represented by a graph, where a constant mass is associated with all vertices, the fundamental mass flows are those over the chords of its graph~\cite{de_persis_pressure_2011}.
In \cite{machado_modeling_2022}, it is shown for a hydraulic network with variable masses associated with certain vertices of its graph representation, as considered in this work, that the fundamental mass flows are those over the chords of a reduced graph of $\mc{G}$, i.e., a graph that contains the same amount of edges but less vertices than $\mc{G}$.
In what follows, let $G$ denote this reduced graph.
To obtain $G$, we merge the vertices from $\mc{V}$ associated with the same TESs.
This yields the reduced graph of $\mc{G}$ representing the same hydraulic network, but assigning a constant mass to all vertices of $G$.
The mass flows over $\mc{F}$ of $G$ are also the fundamental mass flows of the hydraulic network represented by $\mc{G}$ \cite{machado_modeling_2022}.

To describe the relations between the fundamental mass flows and $q_{\mrm{e}}$, let $q_{\mrm{c}} \in \R_+^{|\mc{F}|}$ denote a vector that collects the mass flows over the chords $\mc{F}$.
Then, it is possible to write $q_{\mrm{e}} = F^\top q_{\mrm{c}}$ \cite{machado_modeling_2022} and obtain the ODE-based balance of mass
\begin{equation} \label{eq:reduced mass balance}
    \dot m_{\mrm{h}}(t) = (B)_{v \in \mc{V}_{\mrm{h}}} F^\top q_{\mrm{c}}(t).
\end{equation}

\subsection*{Balance of energy}
From Assumption \ref{ass1}, it follows that the thermodynamic state of each component of the DHG is characterized by its temperature. We assign an average temperature to the respective control volume of each pipe, HX, pipe intersection and TES layer. 
Thus, any vertex $v \in \mc{V}$ or any edge $e \in \mc{E}$ also has an average temperature.
Let ${T_{\mrm{v}}\in\R^{|\mc{V}|}}$ and ${T_{\mrm{e}}\in\R^{|\mc{E}|}}$ denote vectors collecting the vertices' temperatures and the edges' temperatures, respectively.
Following \cite{krug_nonlinear_2021, machado_modeling_2022}, we assume the incoming temperature of an edge to be equal to the temperature of its source vertex as well as the outgoing temperature of an edge to be equal to its average temperature.
\begin{comment}
Furthermore, we introduce the following coupling conditions between the thermal state of neighboring components of the DHG \cite{krug_nonlinear_2021}.
\begin{assumption} \label{ass: coupling equations}
    (i) The incoming temperature of an edge equals the temperature of its source vertex.
    (ii) The outgoing temperature of an edge equals its average temperature.
\end{assumption}
To model this relation, we collect all outgoing and incoming edges of vertex $v_l \in \mc{V}$ in the sets $\mc{O}_l = \{ e \in \mc{E} ~:~ e = (v_l,w) \}$ and $\mc{I}_l = \{ e \in \mc{E} ~:~ e = (w,v_l)$, respectively.
Then, from Assumption \ref{ass: coupling equations} it follows that $(T_{\mrm{e}}(0,t))_{e \in \mc{O}_l} := (T_{\mrm{v}}(t))_l$ for $v_l \in \mc{V}$ and $(T_{\mrm{e}}(L_{\mrm{e,}i},t))_{e \in \mc{O}_l} := (T_{\mrm{e}}(t))_{e \in \mc{O}_l}$ hold, where $T_{\mrm{e}} \in \R^{|\mc{E}|}$ and $T_{\mrm{v}} \in \R^{|\mc{V}|}$ denote vectors collecting temperatures edges and vertices, respectively.  
\end{comment}
Additionally, we define the mass matrices $M_{\mrm{v}} = \diag{m_{\mrm{v}}}$ and $M_{\mrm{e}} = \diag{m_{\mrm{e}}}$ collecting the mass of vertices and edges on their diagonals, respectively, and introduce the abbreviation $B_+ = \frac{1}{2}(|B|+B)$ and $B_- = \frac{1}{2}(|B|-B)$. Then, it is possible to write the balance of energy at vertices and edges in vector form as follows \cite{machado_modeling_2022}: 
\begin{equation} \label{eq:energy balance reduced form}
    \begin{aligned}
        M_{\mrm{v}}(t) \dot T_{\mrm{v}}(t) = &-\diag{B_+ F^\top q_{\mrm{c}}(t) + \kappa_{\mrm{v}}} T_{\mrm{v}}(t) \\
        &+ B_+ \diag{F^\top q_{\mrm{c}}(t)} T_{\mrm{e}}(t) + \kappa_{\mrm{v}} T_{\mrm{a}}, \\
        M_{\mrm{e}} \dot T_{\mrm{e}}(t) =\, &\diag{F^\top q_{\mrm{c}}(t)} B_-^\top T_{\mrm{v}}(t) \\
        &- \diag{F^\top q_{\mrm{c}}(t) + \kappa_{\mrm{e}}} T_{\mrm{e}}(t) + \kappa_{\mrm{e}} T_{\mrm{a}} \\
        &+ P_{\mrm{pr}}(t) - P_{\mrm{d}}(t),
    \end{aligned}
\end{equation}
where $\kappa_{\mrm{v}} \in \R_+^{|\mc{V}|}$ and $\kappa_{\mrm{e}} \in \R_+^{|\mc{E}|}$ denote vectors collecting heat loss coefficients of vertices and edges, respectively, and ${T_{\mrm{a}} \in \R}$, ${P_{\mrm{pr}} \in \R_+^{|\mc{E}|}}$ and ${P_{\mrm{d}} \in \R_+^{|\mc{E}|}}$ denote ambient temperature, injected and extracted heat flow at the edges, respectively.

\subsection*{State space model}
To conclude the thermo-hydraulic modeling, let us summarize equations \eqref{eq:reduced mass balance} and \eqref{eq:energy balance reduced form} into an ODE-based state space model of the form
\begin{equation} \label{eq:overall state model}
    \begin{aligned}
        \dot x(t) &= f(x,u,d) =  M(t)^{-1} \big( A(t) x(t) + E_{\mrm{u}} u(t) + E_{\mrm{d}} d(t) \big), 
    \end{aligned}
\end{equation}
using the state vector $x=\begin{bmatrix} m_{\mrm{h}}^\top & T_{\mrm{v}}^\top & T_{\mrm{e}}^\top\end{bmatrix}^\top \in \R^n$, the control input vector $u=\begin{bmatrix} q_{\mrm{c}}^\top & P_{\mrm{pr}}^\top\end{bmatrix}^\top \in \R^m$, and the disturbance vector $d=\begin{bmatrix} P_{\mrm{d}}^\top & T_{\mrm{a}} \end{bmatrix}^\top \in \R^p$, 
%\begin{equation*}
%    \begin{aligned}
%        x&=\begin{bmatrix} m_{\mrm{h}}^\top & T_{\mrm{v}}^\top & T_{\mrm{e}}^\top\end{bmatrix}^\top \in \R^n,~
%        u=\begin{bmatrix} q_{\mrm{c}}^\top & P_{\mrm{pr}}^\top\end{bmatrix}^\top \in \R^m, \\
%        d&=\begin{bmatrix} P_{\mrm{d}}^\top & T_{\mrm{a}} \end{bmatrix}^\top \in \R^p,
%    \end{aligned}
%\end{equation*}
%respectively, 
where $n=|\mc{V}_{\mrm{h}}|+|\mc{V}|+|\mc{E}|$, $m=|\mc{F}|+|\mc{E}_{\mrm{pr}}|$ and $p=|\mc{E}_{\mrm{d}}|+1$.
Then, we define the matrices
\begin{align}
   % \begin{aligned}
    \label{eq:def_M_A}
         M(t)^{-1}&= \begin{bmatrix} I_{|\mc{V}_{\mrm{h}}|} & 0 & 0 \\ 0 & M_{\mrm{v}}(t) & 0 \\ 0 & 0 & M_{\mrm{e}} \end{bmatrix}^{-1},\; 
         \end{align}
    \begin{align*}
    A(t) &= \begin{bmatrix}
        0 & 0 \\ 
        0 & \Tilde{A}(q_{\mrm{c}}(t))-\tilde A_0
        \end{bmatrix},~ \tilde A_0={\rm diag}\left(\begin{bmatrix}
            \kappa_{\mrm{v}} \\ \kappa_{\mrm{e}}
        \end{bmatrix}\right)
   % \end{aligned}
\end{align*}
\begin{equation}
\label{eq:def_tilde_A}
    \Tilde{A}(q_{\mrm{c}})=\begin{bmatrix}
        -\diag{B_+ F^\top q_{\mrm{c}} } & B_+ \diag{F^\top q_{\mrm{c}}} \\
        \diag{F^\top q_{\mrm{c}}} B_-^\top & -\diag{F^\top q_{\mrm{c}} }
    \end{bmatrix},
\end{equation}
\begin{equation*}
    \begin{aligned}
        E_{\mrm{u}} = \begin{bmatrix}
        (B)_{v \in \mc{V}_{\mrm{h}}} F^\top & 0 \\
        0 & 0 \\
        0 & I_{|\mc{E}|}
    \end{bmatrix},&~
        E_{\mrm{d}} = \begin{bmatrix}
        0 & 0 \\
        0 & \kappa_{\mrm{v}} \\
        -I_{|\mc{E}|} & \kappa_{\mrm{e}}
    \end{bmatrix},
    \end{aligned}
\end{equation*}
to write the dynamics \eqref{eq:reduced mass balance} and \eqref{eq:energy balance reduced form} in the form of \eqref{eq:overall state model}.

For the design of the MPC, we consider an explicit Euler discretization of the system dynamics \eqref{eq:overall state model} with step size ${\Delta t\in\R_+}$ given by 
\begin{equation} \label{eq:euler}
    \begin{aligned}
        x(k+1) &= f_\delta(x(k), u(k), d(k)) \\
        &:=x(k) + \Delta t \, f(x(k), u(k), d(k)),
    \end{aligned}
\end{equation}
where the discrete-time input $u(k)$ is given by the continuous time input at time $k\,\Delta t$ for $k \geq 0$.

\section{Model predictive controller} \label{sec: MPC}
We follow \cite{chen_quasi-infinite_1998, Rajhans2017} and use a quadratic stage cost function, a terminal cost function and a terminal region defined as
\begin{equation}\label{eq:terminal_ingredients}
    \begin{aligned}
        \ell(x,u) &= \|x-\overline{x}\|_Q^2 + \|u-\overline{u}\|_R^2,~
        C_P(x) = \|x - \overline{x}\|_P^2, \\
        \mbb{X}_{C_P,\alpha} &= \{x \in \R^n ~:~ C_P(x) \leq \alpha\}, \text{ for some }\alpha > 0,
    \end{aligned}
\end{equation}
respectively, where $Q \in \R^{n \times n}$, $R \in \R^{m \times m}$, $P \in \R^{n \times n}$ denote positive definite weight matrices and ${(\overline{x}, \overline{u}, \overline{d})}$ denote a steady state tuple of \eqref{eq:euler}.
Moreover, we collect lower and upper bounds for the states as well as for the control inputs in the vectors $x_{\mrm{lb}} \in \R^n$ and $x_{\mrm{ub}} \in \R^n$ as well as  $u_{\mrm{lb}} \in \R^m$ and $u_{\mrm{ub}} \in \R^m$, respectively.
The vector $q_{\mrm{ub}} \in \R_+^{|\mc{E}|}$ collects the upper limits for the mass flow over each pipe.
Then, we can define a set describing the state constrains and a set describing the control input constraints as follows
\begin{equation*}
    \begin{aligned}
        \mbb{X} &= \{ x \in \R^n~:~ x_{\mrm{lb}} \leq x \leq x_{\mrm{ub}} \}, \\
        \mbb{U} &= \{ u \in \R^m~:~ F^\top q_{\mrm{c}} \leq q_{\mrm{ub}}, u_{\mrm{lb}} \leq u \leq u_{\mrm{ub}} \},
    \end{aligned}
\end{equation*}
respectively.
Now, we can formulate the corresponding MPC problem with terminal ingredients for DHGs.
\begin{problem}[MPC with terminal ingredients for DHG]\label{prob:NMPC for DHG}
Consider the discrete-time dynamics \eqref{eq:euler}. At time $k_0$, given the initial state $x_{k_0} \in \mbb{X}$, and a~desired steady state tuple $(\overline{x},\overline{u},\overline{d})$, find a state and control input trajectory $(x_{k_0}^*(\cdot),u_{k_0}^*(\cdot))$ that solves
\begin{equation*}
    \begin{aligned}
        \underset{x(\cdot),u(\cdot)}{\min} ~ & \sum_{k=0}^{N-1} \ell(x(k),u(k)) + C_P(x(N)) \\
        \text{s.t. } & x(k+1) = f_\delta(x(k),u(k),\overline{d}), \quad k=0,\ldots,N-1,\\
        & (x(k), u(k)) \in \mbb{X} \times \mbb{U}, \quad k=0,\ldots,N-1,\\
        & x(N) \in \mbb{X}_{C_P,\alpha}, \\
        & x(0) = x_{k_0}.
    \end{aligned}
\end{equation*}
\end{problem}
Note that knowledge of future disturbance values, i.e., perfect forecast, needs to be assumed in order to solve Problem \ref{prob:NMPC for DHG}.
By iteratively solving Problem~\ref{prob:NMPC for DHG}, we obtain an optimal input sequence $u_{k_0}^*=(u_{k_0}^*(0),u_{k_0}^*(1),\ldots,u_{k_0}^*(N-1))\in \mathbb{U}^{N}$.
The MPC closed-loop control law $\mu:\mbb{X} \rightarrow \mbb{U}$ is then obtained by using only the first element of the input sequence $u_{k}^*(0)$ for the given initial state $x_{k}$, i.e., $\mu(x_{k}):=u_{k}^*(0)$.
Applying $\mu(x_{k_0})$ will result in a system state $x(k_0+1)=x_{k_0+1}$, the time-horizon is moved one step ahead and Problem~\ref{prob:NMPC for DHG} is solved again with initial value $x_{k_0+1}$ instead of $x_{k_0}$.
%The closed-loop solution resulting from $u\mapsto\mu(x)$ is denoted by $x_{\mu}(k)$.
\vspace{-1.5mm}
\subsection*{Asymptotic stabilization of set points}
In the following, we show that the states of the system \eqref{eq:euler} will asymptotically converge to a given desired steady state tuple $(\overline{x},\overline{u},\overline{d})$, when the control law $\mu(\cdot)$ is applied.
Recall that for a discrete time system $x(k+1)=f_\delta(x(k))$, $x(0)=x_0$ an equilibrium $\overline{x}\in\mbb{X}$ is called asymptotically stable if for each $\varepsilon>0$ there exists $\eta(\varepsilon)>0$ such that $\|x_{0}-\overline{x}\|<\eta(\varepsilon)$ implies $\|x(k)-\overline{x}\|<\varepsilon$ for all $k\geq 0$ and that  $x(k)\rightarrow\overline{x}$ holds as $k\rightarrow\infty$. 
Furthermore, we find it useful to recall the meaning of stabilizability.
\begin{defi}[Stabilizability \cite{Heij2021}] \label{def: stabilizability}
    The linear discrete-time system $x(k+1)=A_\delta x(k) + B_\delta u(k)$ is called stabilizable if there exists a matrix $G$, such that $A_\delta + B_\delta G$ is stable, i.e., all its eigenvalues lie in the unit disc.
\end{defi}

It was shown in \cite[Theorem 1]{Rajhans2017} that the closed-loop system's equilibrium is asymptotically stable, see also \cite{chen_quasi-infinite_1998}, if the following conditions hold.
\begin{condition} \label{ass: mpc standard}
    (i)~$f_\delta$ is continuously differentiable. 
    (ii)~$f_\delta(\overline{x},\overline{u},\overline{d})=\overline{x}$.
    (iii)~The discrete time system has a unique solution for every input sequence and any initial value.
    (iv)~$\mbb{U}$ is compact and $\overline{u}$ lies in the interior of $\mbb{U}$.
    (v)~The states are perfectly measured.
    (vi)~The MPC algorithm is \emph{initially feasible}, i.e., Problem~\ref{prob:NMPC for DHG} has a solution for the initial state $x_{k_0}$, (vii) the linearized system of \eqref{eq:euler}, i.e.,
    \begin{equation*}
        x(k+1)= \overline{x} + \left.\frac{\partial f_\delta}{\partial x}\right\rvert_{(\overline{x},\overline{u},\overline{d})} (x(k)-\overline{x}) + \left.\frac{\partial f_\delta}{\partial u}\right\rvert_{(\overline{x},\overline{u},\overline{d})}(u(k)-\overline{u}),
    \end{equation*} is stabilizable in the sense of Definition \ref{def: stabilizability}.
\end{condition}
The foregoing Conditions \ref{ass: mpc standard}.(i) to \ref{ass: mpc standard}.(v) are trivially fulfilled.
Condition \ref{ass: mpc standard}.(vi) is usually needed in the literature, see e.g. \cite{GrunePannek17}, and cannot be proven in general. 

We will show that Condition \ref{ass: mpc standard}.(vii) is satisfied for any DHG that can be described by the system \eqref{eq:euler}. 
For this, we need the following technical assumption, which can be shown to hold due to the hydraulic decoupling introduced by TESs in typical DHG configurations and in particular holds for the exemplary DHG from Figure \ref{fig:exmplDHG_skizze} and \ref{fig:exmplDHG_graph}.
\begin{assumption}
For \eqref{eq:euler}, $\ker F(B)_{v\in\mc{V}_{\mrm{h}}}^\top=\{0\}$ holds.
\label{ass:FB}
\end{assumption}
Furthermore, recall \cite[Lemma~2]{machado_modeling_2022}.
\begin{lemma}
 \label{lem:juan}
    For the steady state mass flows $\overline{q}_{\mrm{c}}$, the matrix $\tilde A$ given by \eqref{eq:def_tilde_A} fulfills $\tilde A(\overline{q}_{\mrm{c}})+\tilde A(\overline{q}_{\mrm{c}})^\top\leq 0$.
\end{lemma}
\begin{prop} \label{prop: stabilizability}
Consider the discretized system \eqref{eq:euler} with step size $\Delta t>0$ and Assumption~\ref{ass:FB}. 
%If $\ker F(B)_{v\in\mathcal{V}_h}^\top=\{0\}$ and if $\Delta t$ is sufficiently small, then Condition \ref{ass: mpc standard} holds.
There exists a sufficiently small $\Delta t$, such that Condition \ref{ass: mpc standard} holds.
\end{prop}
\begin{proof}
We prove that Condition \ref{ass: mpc standard}.(vii) is satisfied, by constructing a matrix $G$, such that Definition \ref{def: stabilizability} holds for the system \eqref{eq:euler}.
%Using Definition \ref{def: stabilizability}, we can prove that Condition \ref{ass: mpc standard}.(vii) holds by finding a stabilizing exemplary control law for \eqref{eq:euler}.
We define the shifted variables $\tilde x(k):=x(k)-\overline{x}$ and $\tilde u(k):=u(k)-\overline{u}$.
Then, the linearized discrete time system from Condition \ref{ass: mpc standard}.(vii) can be rewritten using \eqref{eq:euler} as
\begin{align}
x(k+1) \nonumber &=\overline{x} + \frac{\partial f_\delta}{\partial x}\tilde x(k)+ \frac{\partial f_\delta}{\partial u}\tilde u(k)\nonumber \\
&=x(k)+\Delta t \left(\frac{\partial f}{\partial x}\tilde x(k)+ \frac{\partial f}{\partial u}\tilde u(k)\right),\label{eq:discrete_with_cnts}
\end{align}
where the derivatives are evaluated at $(\overline{x},\overline{u},\overline{d})$. 
To compute the derivatives, we define ${A_0 = -\diag{\begin{bmatrix} 0 & \kappa_v^\top & \kappa_e^\top \end{bmatrix}^\top}}$, ${e_i\in\mathbb{R}^{|\mathcal{F}|}}$ as the ${i^{\mrm{th}}}$ canonical unit vector, i.e., ${(e_i)_j=1}$ if ${i=j}$, and zero entries otherwise, and ${A_i=\begin{bmatrix} 0\\ \hat A_i \end{bmatrix}}$, where ${\hat A_i=\tilde A(e_i)}$, is given by \eqref{eq:def_tilde_A}. % and $e_i\in\mathbb{R}^{|\mathcal{F}|}$ is the $i^{\mrm{th}}$ canonical unit vector, i.e., $(e_i)_j=1$ if $i=j$, and zero entries otherwise.
Then, we can write %the right-hand side $f$ in the following way
\begin{equation*}
f(x,u,\overline{d})=M^{-1}(A_0+\sum_{i=1}^{|\mathcal{F}|} u_i A_i)x + E_u u + E_d \overline{d}.
\end{equation*}
%where $A_0 = -\diag{0,\kappa_v,\kappa_e}$, and 
%obtained by decomposing the matrix $A$ in \eqref{eq:overall state model} and which are for $i=1,\ldots,|\mathcal{F}|$ of the form 
%$A_i=\begin{bmatrix}
%    0\\ \hat A_i
%\end{bmatrix}$,
%where $\hat A_i=\tilde A(e_i)$, is given by \eqref{eq:def_tilde_A} and $e_i\in\mathbb{R}^{|\mathcal{F}|}$ is the $i^{\mrm{th}}$ canonical unit vector, i.e., $(e_i)_j=1$ if $i=j$, and zero entries otherwise.
%$A_i=\begin{bmatrix}
 %   0\\ \hat A_i
%\end{bmatrix}$ \textcolor{red}{where $\hat A_i\in\mathbb{R}^{(n-|\mathcal{V}_{\mrm{h}}|)\times n}$ is given by 
%\[
%...
%\]
%and $A_i=0$ for $i=|\mathcal{F}|+1,\ldots,m$.}
Using the steady state condition $f(\overline{x},\overline{u},\overline{d})=0$, abbreviating $\hat A(\overline{x}):=\begin{bmatrix}\hat A_1\overline{x} \ldots  \hat A_{|\mathcal{F}|}\overline{x}\end{bmatrix}$ and by considering the positive definite matrix $\overline{M}$, which is obtained from $M(t)$ given in \eqref{eq:def_M_A} by using the steady state values for the vertex masses, this leads to 
\begin{align}
\frac{\partial f}{\partial x}(\overline{x},\overline{u},\overline{d})&=\overline{M}^{-1}(A_0+\sum_{i=1}^{|\mathcal{F}|} \overline{u}_iA_i),\label{eq:jacobi_f_x}\\
\frac{\partial f}{\partial u}(\overline{x},\overline{u},\overline{d})&=\overline{M}^{-1}\left(\begin{bmatrix}A_1\overline{x} \ldots A_{|\mathcal{F}|}\overline{x}&0\end{bmatrix}+E_u\right)\nonumber \\ &=\overline{M}^{-1}\begin{bmatrix}
    (B)_{v\in\mathcal{V}_h}F^\top &0\\  \hat A(\overline{x}) & \begin{bmatrix}
        0\\I
    \end{bmatrix}
\end{bmatrix}. \label{eq:jacobi_f_u}
\end{align}

Furthermore, since $\ker F(B)_{v\in\mathcal{V}_h}^\top=\{0\}$ holds by Assumption~\ref{ass:FB}, the right inverse $W$ of $(B)_{v\in\mathcal{V}_h}F^\top$ exists, i.e., $(B)_{v\in\mathcal{V}_h}F^\top W=I$.
Then, for some $\varepsilon>0$, we define the following state feedback  
\begin{align}
\label{eq:feedback}
\tilde u(k) &=\begin{bmatrix}
 q_c-\overline{q}_c\\ P_{pr}-\overline{P}_{pr}\end{bmatrix}=G \tilde x(k)\\ &=\begin{bmatrix}
   -\varepsilon W &\varepsilon W(\hat A(\overline{x})W)^\top \\ 0&0
\end{bmatrix}\begin{bmatrix}
m_{\mrm{h}}(k)-\overline{m}_{\mrm{h}} \\ \begin{bmatrix}
    T_v(k)-\overline{T_v}\\ T_e(k)-\overline{T_e}
\end{bmatrix}\end{bmatrix}. \nonumber %+v_1
\end{align}
Using \eqref{eq:discrete_with_cnts} together with \eqref{eq:jacobi_f_x}, \eqref{eq:jacobi_f_u} and \eqref{eq:feedback}, leads to  
\[
\tilde x(k+1)=\tilde x(k)+\Delta t\overline{M}^{-1}\begin{bmatrix}
    K_1 &-K_2^\top \\ K_2 & K_3
\end{bmatrix} \tilde{x}(k)=: A_d\tilde x(k),
\]
with $K_1:=-\varepsilon I$, $K_2:= -\varepsilon\hat A(\overline{x}) W$ as well as
\begin{align*}    
K_3:=-\sum_{i=1}^{|\mathcal{F}|}\overline{u}_i\hat A_i
    -\tilde A_0+\varepsilon (\hat A(\overline{x}))W(\hat A(\overline{x})W)^\top.
\end{align*}
Then, $K_1$ is negative definite and by choosing $\varepsilon$ sufficiently small, it follows from \cite[Lemma 11]{Schiffer17} that $K_3$ fulfills $z^\top( K_3+K_3^\top)z\leq 0$. Consider now 
\begin{align}
\label{eq:discrete_lyap}
&~~~~A_d^\top \overline{M}A_d-\overline{M} \\ &=2\Delta t\begin{bmatrix}
    K_1&0\\0& K_3
\end{bmatrix} + \Delta t^2\begin{bmatrix}
    K_1&K_2^\top\\-K_2& K_3
\end{bmatrix}\overline{M}^{-1}\begin{bmatrix}
    K_1&-K_2^\top\\K_2& K_3
\end{bmatrix}. \nonumber
\end{align}
If $\Delta t>0$ is sufficiently small, then it follows again from \cite[Lemma 11]{Schiffer17}, that \eqref{eq:discrete_lyap} is negative definite  
and therefore the closed-loop system $\tilde x(k+1)=A_d\tilde x(k)$ has the Lyapunov function ${V(\tilde x)=\tilde x^\top\overline{M}\tilde x}$ with the discrete derivative along the solutions ${\Delta V(\tilde x(k))=\tilde x(k)^\top(A_d^\top \overline{M}A_d-\overline{M})\tilde x(k)}$. % \begin{align*}\Delta V(\tilde x(k))&=V(\tilde x(k+1))-V(\tilde x(k))\\&=\tilde x(k)^\top(A_d^\top \overline{M}A_d-\overline{M})\tilde x(k).\end{align*}
% as a solution for the discrete-time Lyapunov inequality.
Hence, the feedback \eqref{eq:feedback} stabilizes the linearization of \eqref{eq:euler} in $(\overline{x},\overline{u},\overline{d})$. Therefore, Condition~\ref{ass: mpc standard} is fulfilled.
\end{proof}
Thus, if the DHG fulfills $\ker F(B)_{v\in\mathcal{V}_h}^\top=\{0\}$ and if the step size $\Delta t$ is sufficiently small, then the closed-loop system's equilibrium resulting from applying the control law $\mu_{N}(\cdot)$ is asymptotically stable.
%\begin{remark}
%    We use the assumption that $\ker F(B)_{v\in\mathcal{V}_h}^\top=\{0\}$ holds to proof Proposition \ref{prop: stabilizability}. \TODO{Interpretation necessary: What does that mean for DHG?}
%\end{remark}

\section{Case Study} \label{sec: case study}
%%%%%%%%%%%%%%%%%%%%%%%%%
%%%%%%%%%%%%%%%%%%%%%%%%%
%%%%%%%%%%%%%%%%%%%%%%%%%
In this section, we demonstrate the practicability of the MPC approach outlined in Section ~\ref{sec: MPC} via a numerical case study.
For this, the exemplary DHG shown in Figure~\ref{fig:exmplDHG_skizze} is modeled as described in Section~\ref{sec: model} and numerically implemented\footnote{The source code can be accessed in the following repository: \url{https://github.com/max65945/mpcofdhg_ecc24}.} using julia programming language \cite{bezanson_julia_2012}.
In what follows, we first describe the parametrization of the DHG model and define nominal operating conditions.
Based on that, we define two steady states with suitable terminal ingredients for each steady state.
Finally, we simulate the stabilization of the two setpoints via MPC with terminal ingredients.

%\mr{
%We set the density of water, the ambient temperature, the specific heat capacity of water, the pipe friction coefficient and the heat loss coefficient to  $\rho = 988.05 \frac{\text{kg}}{\text{m}^3}$, $T_{\mrm{a}} = 10.0^\circ\text{C}$, $c_p = 4.18 \frac{\text{kJ}}{\text{kg}\cdot\text{K}}$, $\lambda = 0.02$ and $(\kappa_{\mrm{v}})_v = (\kappa_{\mrm{e}})_e = 0.2 \frac{\text{kJ}}{\text{K}\cdot\text{s}}$, respectively.
%As indicated in Figure \ref{fig:exmplDHG_skizze}, three temperature levels evolve during operation. 
%The nominal operating temperature for the low temperature level (blue), the medium temperature level (orange) and the high temperature level (red) is set to $T_{\mrm{low}}=20^\circ\text{C}$, $T_{\mrm{mid}}=45^\circ\text{C}$ and $T_{\mrm{high}}=75^\circ\text{C}$ respectively.
%Additionally, we define a nominal heat flow demand of ${P_{\mrm{d,}1}^{\mrm{nom}} = 1.5 \text{MW}}$, ${P_{\mrm{d,}2}^{\mrm{nom}} = 2.5 \text{MW}}$ and ${P_{\mrm{d,}3}^{\mrm{nom}} = 1 \text{MW}}$ and set the length of each pipe to $L = 500 \text{m}$.
%Using the Darcy-Weisbach equation, the pipe diameters are designed such that the specific pressure drop equals $300~\text{Pa}/\text{m}$ in nominal operation.
%}
The length of each edge is set to $L = 500 \text{m}$, the total masses of the TESs are set to $m_{\mrm{tes,}1} = 50 \cdot 10^3\text{kg}$ and $m_{\mrm{tes,}2} = 30 \cdot 10^3\text{kg}$, respectively.%, and the masses of pipe intersections to $(m_{\mrm{v}})_v = 4000\text{kg}$ for $v \in \mc{E}_{\mrm{p}}$.
The heat loss coefficient is set to $(\kappa_{\mrm{v}})_v = (\kappa_{\mrm{e}})_e = 0.2 \frac{\text{kJ}}{\text{K}\cdot\text{s}}$ for $v \in \mc{V}$, $e \in \mc{E}$. 
Furthermore, we set the density of water, the ambient temperature, the specific heat capacity of water and the pipe friction coefficient to  $\rho = 988.05 \frac{\text{kg}}{\text{m}^3}$, $T_{\mrm{a}} = 10.0^\circ\text{C}$, $c_p = 4.18 \frac{\text{kJ}}{\text{kg}\cdot\text{K}}$, $\lambda = 0.02$, respectively.

We define two steady state set points $(\overline{x}^{I},\overline{u}^{I})$ and $(\overline{x}^{II},\overline{u}^{II})$ for the the DHG dynamics~\eqref{eq:euler}.
In what follows, we use the superscript ($I$ or $II$) to associate the variables
to each of the set points.
One set point is computed for ${P_{\mrm{d,}1}^{I} = 1.5 \text{MW}}$, ${P_{\mrm{d,}2}^{I} = 2.5 \text{MW}}$ and ${P_{\mrm{d,}3}^{I} = 1 \text{MW}}$ with nominal temperatures at vertex $v_1$ and $v_6$ of ${T_{\mrm{v,}1}^{\mrm{nom,}I} = 45 ^\circ\text{C}}$ and ${T_{\mrm{v,}6}^{\mrm{nom,}I} = 75 ^\circ\text{C}}$, respectively.
The pipe diameters are designed using the Darcy-Weisbach equation such that the pressure drop equals $300~\text{Pa}/\text{m}$ when operated at $(\overline{x}^{I},\overline{u}^{I})$.
The second set point is computed for ${P_{\mrm{d,}1}^{II} = 1.5 \text{MW}}$, ${P_{\mrm{d,}2}^{II} = 2 \text{MW}}$ and ${P_{\mrm{d,}3}^{II} = 1 \text{MW}}$ with nominal temperatures at vertex $v_1$ and $v_6$ of ${T_{\mrm{v,}1}^{\mrm{nom,}II} = 46 ^\circ\text{C}}$ and ${T_{\mrm{v,}6}^{\mrm{nom,}II} = 77 ^\circ\text{C}}$, respectively.
With the given parameters, the heat losses are 9\% and 8\% in the stationary states $(\overline{x}^{I},\overline{u}^{I})$ and $(\overline{x}^{II},\overline{u}^{II})$, respectively.

Note that the Conditions~\ref{ass: mpc standard}~(i)-(vi) are satisfied for the described setup and the steady states $(\overline{x}^{I},\overline{u}^{I})$ and $(\overline{x}^{II},\overline{u}^{II})$.
Furthermore, Assumption~\ref{ass:FB} holds, and it can be verified that Proposition~\ref{prop: stabilizability} is applicable for $\Delta t = 60\text{s}$.
Therefore, the Condition~\ref{ass: mpc standard}~(vii) holds, which means that the exemplary DHG is stabilizable in the sense of Definition \ref{def: stabilizability}.
Hence, the linear quadratic regulator (LQR) can be used to derive suitable terminal ingredients \cite{chen_quasi-infinite_1998}.
%We use 
%\begin{equation*}
%    \begin{aligned}
%        Q &= \begin{bmatrix} I_{|\mc{V}_{\mrm{h}}|} \cdot m_{\mrm{n}}^{-1} & 0 \\ 0 & I_{|\mc{V}|+|\mc{E}|} \cdot T_{\mrm{n}}^{-1} \end{bmatrix}, \\
%        R &= \begin{bmatrix} I_{|\mc{F}|} \cdot q_{\mrm{n}}^{-1} & 0 \\ 0 & I_{|\mc{E}_{\mrm{pr}}|} \cdot P_{\mrm{n}}^{-1} \end{bmatrix},
%    \end{aligned}
%\end{equation*}
%where $m_{\mrm{n}} = 40 \cdot 10^3 \text{kg}$, $T_{\mrm{n}} = 40 ^\circ\text{C}$, $q_{\mrm{n}} = 20 \text{kg} / \text{s}$ and $P_{\mrm{n}} = 5 \text{MW}$ denote nominal values for all the TESs masses, DHG temperatures, pipe mass flows and injected heat flows, respectively, as weight matrices to obtain the LQR feedback law.
The prediction horizon is set to ${N = 60}$.
Then, terminal ingredients can be derived via an auxiliary nonlinear optimization problem \cite{chen_quasi-infinite_1998}.
Without going into the details of this optimization problem, we point out that due to the nonlinearity of the DHG model, it is not possible to make a statement about the globality of the optimizer without further investigation, which is beyond the scope of this paper.
Instead, we solve the problem iteratively with 100 different normally distributed initial values and assume that in this way we can approximate the global solution of the optimizer with sufficient accuracy.

The procedure results in one ellipsoid per set point.
Each ellipsoid describing a terminal region (and a terminal cost) has dimension 18 and is defined by the solution $P^{(\cdot)}$ of the Lyapunov equation for the linearized closed-loop system.
We denote each solution by $P^{I}$ and $P^{II}$, with corresponding $\alpha^{I}$ and $\alpha^{II}$, respectively; see~\eqref{eq:terminal_ingredients}.

In order to give a rough idea of the size of the terminal region, we identify the smallest box containing each ellipsoid and compute its one-dimensional projection on the coordinates corresponding to the states of the TESs.
Here, we select all states associated with the TESs as states of particular interest because solely the TESs allow for a variable mass ratio of hot to cold water and have the largest inertia within the DHG due to their large total masses.
In Table~\ref{tab: terminal region approx.}, we display the distance with respect to the associated steady state value representing the projections of the approximate terminal region of TES~1 and TES~2 onto the masses $\Delta m_{\mrm{v,}1} \in \R$, $\Delta m_{\mrm{v,}6} \in \R$ and temperatures $\Delta T_{\mrm{v,}1} \in \R$, $\Delta T_{\mrm{v,}6} \in \R$, respectively.
\begin{table}
    \centering
        \caption{Values approximating the terminal region for states associated with TESs.}
    \label{tab: terminal region approx.}
    \begin{tabular}{c| c| c| c| c}
        \toprule
        & $\Delta m_{\mrm{v,}1}$ & $\Delta T_{\mrm{v,}1}$ & $\Delta m_{\mrm{v,}6}$ & $\Delta T_{\mrm{v,}6}$ \\
        \bottomrule
        I & $0.75 \text{t}$ & $2.05 ^\circ\text{C}$ & $0.45 \text{t}$ & $0.4 ^\circ\text{C}$\\
        II & $0.75 \text{t}$ & $3.3 ^\circ\text{C}$ & $0.45 \text{t}$ & $2.07 ^\circ\text{C}$\\
        \bottomrule
    \end{tabular}
\end{table}

In order to assess the closed-loop system's performance over $N_{\text{sim}} \in \N$ discrete-time steps in the face of a sudden set point change at the discrete-time step $k_{\mrm{step}} \in \N$, we use the calculated set points and define two different controllers, namely MPC 1 and MPC 2.
We use the stage cost, terminal cost and terminal region
\begin{equation*}
    \begin{aligned}
        \ell_1(x,u)&=\begin{cases}
                        \|x-\overline{x}^I\|_Q^2 + \|u-\overline{u}^I\|_R^2,~\text{if}~k \in \mc{I}_1, \\
                        \|x-\overline{x}^{II}\|_Q^2 + \|u-\overline{u}^{II}\|_R^2,~\text{if}~k \in \mc{I}_2,
                    \end{cases} \\
        C_1(x)&=\begin{cases}
                        C_{P^I}(x),~\text{if}~k \in \mc{I}_1, \\
                        C_{P^{II}}(x),~\text{if}~k \in \mc{I}_2,
                    \end{cases} \\
        \mbb{X}_{C,1}&=\begin{cases}
                        \mbb{X}_{C_{P^I},\alpha^I},~\text{if}~k \in \mc{I}_1, \\
                        \mbb{X}_{C_{P^{II}},\alpha^{II}},~\text{if}~k \in \mc{I}_2,
                    \end{cases}
    \end{aligned}
\end{equation*}
respectively, where $\mc{I}_1 = \{0,\ldots,k_{\mrm{step}}-1\} \subset \N$ and $\mc{I}_2 = \{k_{\mrm{step}},N_{\mrm{sim}}\}\subset \N$, to define MPC 1.
Thus, for $k \in \mc{I}_1$, MPC~1 aims to stabilize $(\overline{x}^{I},\overline{u}^{I})$.
Then, for $k \in \mc{I}_2$, $(\overline{x}^{II},\overline{u}^{II})$ is stabilized.
To obtain MPC~2, we use a piece-wise constant reference trajectory
\begin{equation} \label{eq:piece-wise constant reference}
    (\overline{x}(k),\overline{u}(k)) = \begin{cases}
        (\overline{x}^{I},\overline{u}^{I}), \text{ for } k \in \mc{I}_1, \\
        (\overline{x}^{II},\overline{u}^{II}), \text{ for } k \in \mc{I}_2,
    \end{cases}
\end{equation}
to define the stage cost function $\ell_2(k,x,u) = \|x-\overline{x}(k)\|_Q^2 + \|u-\overline{u}(k)\|_R^2$, together with piece-wise constant terminal cost and terminal region
\begin{equation} \label{eq: C2 X_C2}
    \begin{aligned}
        C_2(k,x)&=\begin{cases}
                        C_{P^I}(x),~\text{if}~k \in \mc{I}_3, \\
                        C_{P^{II}}(x),~\text{if}~k \in \mc{I}_4,
                    \end{cases} \\
        \mbb{X}_{C,2}(k)&=\begin{cases}
                        \mbb{X}_{C_{P^I},\alpha^I},~\text{if}~k \in \mc{I}_3, \\
                        \mbb{X}_{C_{P^{II}},\alpha^{II}},~\text{if}~k \in \mc{I}_4,
                    \end{cases}
    \end{aligned}
\end{equation}
respectively, where $\mc{I}_3 = \{0,\ldots,k_{\mrm{step}}-N-1\} \subset \N$ and $\mc{I}_4 = \{k_{\mrm{step}}-N,\ldots,N_{\text{sim}}\} \subset \N$.
It follows, that $C_2(k,x)$ and $\mbb{X}_{C,2}(k)$ suddenly change at time step $k_{\mrm{step}}-N$, whereas $\ell_2(k,x,u)$ changes for $k \in \mc{I}_1 \cup \mc{I}_4$.
Thus, MPC~1 and MPC~2 only differ within the transition phase, i.e., for $k \in \mc{I}_1 \cup \mc{I}_4$.
Through the application of MPC 2, we investigate the ability of MPC in the context of DHG to utilize knowledge of future heat flow extraction and temperature set points, i.e., to incorporate a predictive behavior into the closed-loop DHG, during the transition phase.

\begin{figure}
    \centering
    \begin{tikzpicture}
    \begin{myGroupPlot}
        % 1st subplot
        \nextgroupplot[ylabel={$[\mathrm{t}]$},xtick={0,0.5,...,5}]
        \addplot [color=Mahogany, line legend,stack plots=false] table [x=time, y=x1_sol_1, col sep=comma] {./results/results_QIHMPC_250324.csv};
        \addlegendentry{$m_{v_1}^{\mathrm{MPC}1}$}
        \addplot [color=Aquamarine, dashed, line legend,stack plots=false] table [x=time, y=x2_sol_1, col sep=comma] {./results/results_QIHMPC_250324.csv};
        \addlegendentry{$m_{v_1}^{\mathrm{MPC}2}$}
        \addplot [color=black, dotted, line legend,stack plots=false, line width=0.8pt] table [x=time, y=x_ss_1, col sep=comma] {./results/results_QIHMPC_250324.csv};
        %\addplot [color=black, dashed, line legend,stack plots=false, line width=.5pt] table [x=time, y=x_max_1, col sep=comma] {./results/results_QIHMPC_250324.csv};

         % 2nd subplot
        \nextgroupplot[ylabel={$[\frac{\mathrm{kg}}{\mathrm{s}}]$},xtick={0,0.5,...,5}]
        \addplot [color=NavyBlue, line legend,stack plots=false] table [x=time, y=u1_sol_1, col sep=comma] {./results/results_QIHMPC_250324.csv};
        \addlegendentry{$q_{e_1}^{\mathrm{MPC}1}$}
        \addplot [color=OrangeRed, dashed, line legend,stack plots=false] table [x=time, y=u1_sol_2, col sep=comma] {./results/results_QIHMPC_250324.csv};
        \addlegendentry{$q_{e_2}^{\mathrm{MPC}1}$}
        \addplot [color=black, dotted, line legend,stack plots=false, line width=0.8pt] table [x=time, y=u_ss_1, col sep=comma] {./results/results_QIHMPC_250324.csv};

        % 3rd subplot
        \nextgroupplot[ylabel={$[\frac{\mathrm{kg}}{\mathrm{s}}]$},xtick={0,0.5,...,5}]
        \addplot [color=Emerald, line legend,stack plots=false] table [x=time, y=u2_sol_1, col sep=comma] {./results/results_QIHMPC_250324.csv};
        \addlegendentry{$q_{e_1}^{\mathrm{MPC}2}$}
        \addplot [color=Plum, dashed, line legend,stack plots=false] table [x=time, y=u2_sol_2, col sep=comma] {./results/results_QIHMPC_250324.csv};
        \addlegendentry{$q_{e_2}^{\mathrm{MPC}2}$}
        \addplot [color=black, dotted, line legend,stack plots=false, line width=0.8pt] table [x=time, y=u_ss_1, col sep=comma] {./results/results_QIHMPC_250324.csv};

        % 4th subplot
        \nextgroupplot[ylabel={$[^\circ \mathrm{C}]$},ytick={42,43,44,45},xtick={0,0.5,...,5}]
        \addplot [color=Brown, line legend,stack plots=false] table [x=time, y=x1_sol_3, col sep=comma] {./results/results_QIHMPC_250324.csv};
        \addlegendentry{$T_{v_1}^{\mathrm{MPC}1}$}
        \addplot [color=PineGreen, dashed, line legend,stack plots=false] table [x=time, y=x2_sol_3, col sep=comma] {./results/results_QIHMPC_250324.csv};
        \addlegendentry{$T_{v_1}^{\mathrm{MPC}2}$}
        \addplot [color=black, dotted, line legend,stack plots=false, line width=0.8pt] table [x=time, y=x_ss_3, col sep=comma] {./results/results_QIHMPC_250324.csv};
        %\addplot [color=black, dashed, line legend,stack plots=false, line width=.5pt] table [x=time, y=x_max_3, col sep=comma] {./results/results_QIHMPC_250324.csv};

        % 5th subplot
        \nextgroupplot[ylabel={$[\mathrm{WM}]$},ytick={4.4,4.7,5,5.3},xtick={0,0.5,...,5}]
        \addplot [color=RawSienna, line legend,stack plots=false] table [x=time, y=u1_sol_10, col sep=comma] {./results/results_QIHMPC_250324.csv};
        \addlegendentry{$P_{\mathrm{pr,}1}^{\mathrm{MPC}1}$}
        \addplot [color=YellowOrange, dashed, line legend,stack plots=false] table [x=time, y=u2_sol_10, col sep=comma] {./results/results_QIHMPC_250324.csv};
        \addlegendentry{$P_{\mathrm{pr,}1}^{\mathrm{MPC}2}$}
        \addplot [color=black, dotted, line legend,stack plots=false, line width=0.8pt] table [x=time, y=u_ss_10, col sep=comma] {./results/results_QIHMPC_250324.csv};
        \addplot [color=black, dashed, line legend,stack plots=false, line width=.5pt] table [x=time, y=u_max_10, col sep=comma] {./results/results_QIHMPC_250324.csv};

        % 6th subplot
        \nextgroupplot[ylabel={$[\mathrm{t}]$},ytick={14.5,14.75,15,15.25},xtick={0,0.5,...,5}]
        \addplot [color=myRedA, line legend,stack plots=false] table [x=time, y=x1_sol_2, col sep=comma] {./results/results_QIHMPC_250324.csv};
        \addlegendentry{$m_{v_6}^{\mathrm{MPC}1}$}
        \addplot [color=myBlueA, dashed, line legend,stack plots=false] table [x=time, y=x2_sol_2, col sep=comma] {./results/results_QIHMPC_250324.csv};
        \addlegendentry{$m_{v_6}^{\mathrm{MPC}2}$}
        \addplot [color=black, dotted, line legend,stack plots=false, line width=0.8pt] table [x=time, y=x_ss_2, col sep=comma] {./results/results_QIHMPC_250324.csv};
        %\addplot [color=black, dashed, line legend,stack plots=false, line width=.5pt] table [x=time, y=x_max_2, col sep=comma] {./results/results_QIHMPC_250324.csv};

        % 7th subplot
        \nextgroupplot[ylabel={$[\frac{\mathrm{kg}}{\mathrm{s}}]$},ytick={8,9},xtick={0,0.5,...,5}]
        \addplot [color=Periwinkle, line legend,stack plots=false] table [x=time, y=u1_sol_7, col sep=comma] {./results/results_QIHMPC_250324.csv};
        \addlegendentry{$q_{e_7}^{\mathrm{MPC}1}$}
        \addplot [color=Orange, dashed, line legend,stack plots=false] table [x=time, y=u1_sol_8, col sep=comma] {./results/results_QIHMPC_250324.csv};
        \addlegendentry{$q_{e_8}^{\mathrm{MPC}1}$}
        \addplot [color=black, dotted, line legend,stack plots=false, line width=0.8pt] table [x=time, y=u_ss_7, col sep=comma] {./results/results_QIHMPC_250324.csv};

        % 8th subplot
        \nextgroupplot[ylabel={$[\frac{\mathrm{kg}}{\mathrm{s}}]$},ytick={8,9},xtick={0,0.5,...,5}]
        \addplot [color=Emerald, line legend,stack plots=false] table [x=time, y=u2_sol_7, col sep=comma] {./results/results_QIHMPC_250324.csv};
        \addlegendentry{$q_{e_7}^{\mathrm{MPC}2}$}
        \addplot [color=Plum, dashed, line legend,stack plots=false] table [x=time, y=u2_sol_8, col sep=comma] {./results/results_QIHMPC_250324.csv};
        \addlegendentry{$q_{e_8}^{\mathrm{MPC}2}$}
        \addplot [color=black, dotted, line legend,stack plots=false, line width=0.8pt] table [x=time, y=u_ss_7, col sep=comma] {./results/results_QIHMPC_250324.csv};

        % 9th subplot
        \nextgroupplot[ylabel={$[^\circ \mathrm{C}]$},xtick={0,0.5,...,5}]
        \addplot [color=RubineRed, line legend,stack plots=false] table [x=time, y=x1_sol_8, col sep=comma] {./results/results_QIHMPC_250324.csv};
        \addlegendentry{$T_{v_6}^{\mathrm{MPC}1}$}
        \addplot [color=YellowGreen, dashed, line legend,stack plots=false] table [x=time, y=x2_sol_8, col sep=comma] {./results/results_QIHMPC_250324.csv};
        \addlegendentry{$T_{v_6}^{\mathrm{MPC}2}$}
        \addplot [color=black, dotted, line legend,stack plots=false, line width=0.8pt] table [x=time, y=x_ss_8, col sep=comma] {./results/results_QIHMPC_250324.csv};
        %\addplot [color=black, dashed, line legend,stack plots=false, line width=.5pt] table [x=time, y=x_max_8, col sep=comma] {./results/results_QIHMPC_250324.csv};

        % 10th subplot
        \nextgroupplot[ylabel={$[\mathrm{WM}]$},ytick={1.2,1.3,1.4},xtick={0,0.5,...,5},xticklabels={0,0.5,...,5}, xlabel={time $[\mathrm{h}]$}]
        \addplot [color=BrickRed, line legend,stack plots=false] table [x=time, y=u1_sol_11, col sep=comma] {./results/results_QIHMPC_250324.csv};
        \addlegendentry{$P_{\mathrm{pr,}2}^{\mathrm{MPC}1}$}
        \addplot [color=Peach, dashed, line legend,stack plots=false] table [x=time, y=u2_sol_11, col sep=comma] {./results/results_QIHMPC_250324.csv};
        \addlegendentry{$P_{\mathrm{pr,}2}^{\mathrm{MPC}2}$}
        \addplot [color=black, dotted, line legend,stack plots=false, line width=0.8pt] table [x=time, y=u_ss_11, col sep=comma] {./results/results_QIHMPC_250324.csv};
        \addplot [color=black, dashed, line legend,stack plots=false, line width=.5pt] table [x=time, y=u_max_11, col sep=comma] {./results/results_QIHMPC_250324.csv};
    \end{myGroupPlot}
\end{tikzpicture}
    \caption{Masses and temperatures at TESs with corresponding mass flow rates as well as heat flow injections for MPC 1 and MPC 2. Set points are indicated by the dotted lines.}
    \vspace{-1mm}
    \label{fig:sim_results}
\end{figure}
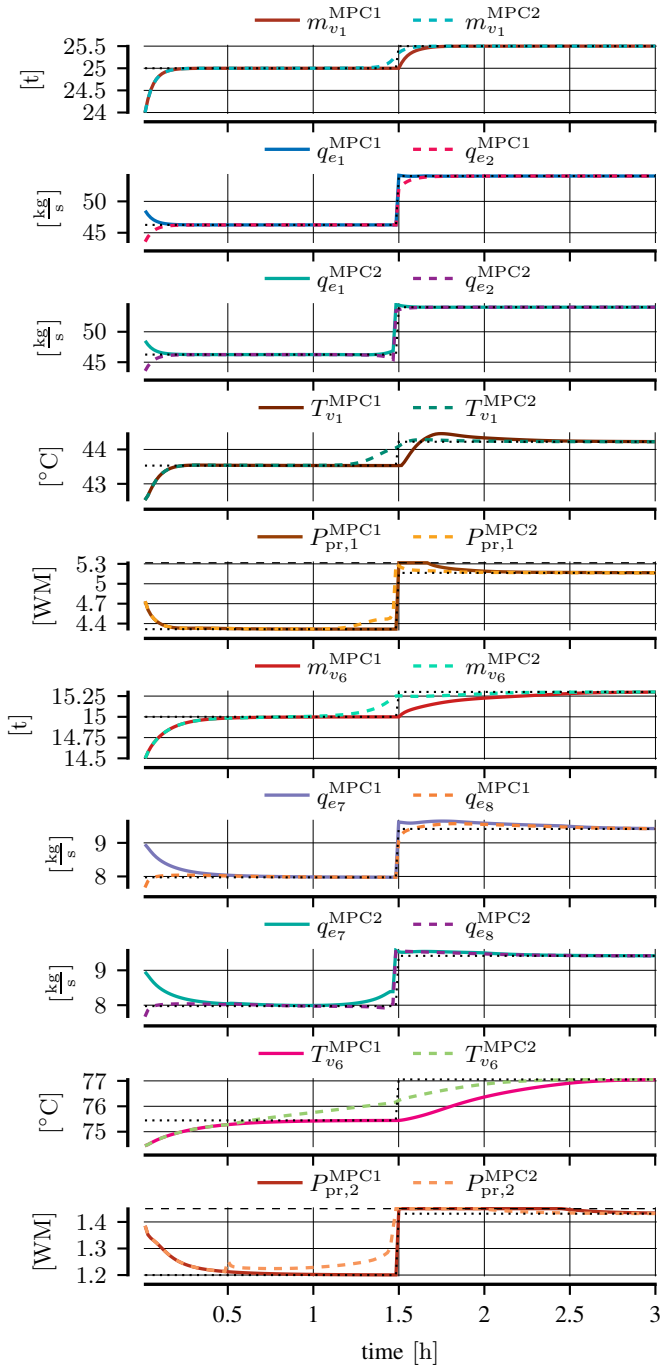

We simulate the closed-loop DHG for $3\text{h}$, i.e., for $N_{\text{sim}} = 180$, and choose $k_{\mrm{step}} = 90$.
The average time required to solve the optimization problem of MPC 1 and MPC 2 is $0.12\text{s}$ and $0.14\text{s}$, respectively.
The maximum time required to solve the optimization problem of MPC 1 and MPC 2 is $0.96\text{s}$ and $0.42\text{s}$, respectively. 
The results of the described simulation are shown in Figure \ref{fig:sim_results}.
It can be observed that the DHG under MPC 1 is steered towards $(\overline{x}^{I},\overline{u}^{I})$ for $k \in \mc{I}_1$ and that all desired set points which are indicated by the black dotted lines are reached before the set point changes.
Afterwards, for $k \in \mc{I}_2$, MPC 1 steers the DHG towards $(\overline{x}^{II},\overline{u}^{II})$ so that all set points are reached latest after $2.75\text{h}$.
This illustrates the theoretical results on asymptotic stabilization of constant set points from Section \ref{sec: MPC}.

For MPC 2, it can be observed that the DHG is steered towards $(\overline{x}^{I},\overline{u}^{I})$ for $k \in \mc{I}_3$.
As soon as the second steady state lies within the prediction horizon of the MPC, i.e., for $k \in \mc{I}_4$, the terminal region and terminal cost of the optimization problem of MPC 2 change according to \eqref{eq: C2 X_C2}.
Additionally, with each iteration for $k \in \mc{I}_1 \cup \mc{I}_4$ the stage cost function is updated according to \eqref{eq:piece-wise constant reference}.
Therefore, we can see that the closed-loop behavior of the DHG already adapts for $k \in \mc{I}_1 \cup \mc{I}_4$, which is manifested in the form of a increase of the temperatures of TES 1 and TES 2 as well as an increase of mass stored in TES 1 and TES 2 towards $(\overline{x}^{II},\overline{u}^{II})$.
%After roughly $1.35\text{h}$, also the water mass stored in each TES is adapted according to the set points $(\overline{x}^{II},\overline{u}^{II})$.
%This change slightly perturbs the TESs temperatures.
%Afterwards, the temperatures of the TESs converge towards $(\overline{x}^{II},\overline{u}^{II})$ and reach the steady state at approximately $t = 2.75 \text{h}$.
Using MPC 2, the steady state $(\overline{x}^{II},\overline{u}^{II})$ is reached after approximately $t = 2.5 \text{h}$.
%Compared to MPC 1, MPC 2 slightly reduces costs about $2.5\%$, but significantly reduces the required control input changes in between successive control actions in the mass flows by up to $4.4$ times and the required heat flows by up to $1.5$ times.
Compared to MPC 1, MPC 2 reduces required control input changes in between successive control actions in the required heat flows by up to $3.3$ times.
This indicates that the incorporation of a stage cost of the form of $l_2(k,x,u)$ is beneficial for smooth control input trajectories.

Additionally, the successful incorporation of control input constraints can be observed for the injected heat flow $P_{\mrm{pr,}1}^{\mrm{MPC}1}$, $P_{\mrm{pr,}2}^{\mrm{MPC}1}$ and $P_{\mrm{pr,}2}^{\mrm{MPC}2}$ for $t \geq 1.5 \text{h}$ since the heat flow injection reaches its limit, but does not overshoot.
Remaining constraints also hold but are never active.

To summarize, this case study demonstrates the practicability of the MPC approach using terminal ingredients for a realistic small scale DHG for two different steady states.
In this process, two MPCs with terminal ingredients are compared that coordinate two heat producers so that three consumers can draw heat from the DHG at desired temperature levels.
Both MPCs stabilize given piece-wise constant set points and satisfy constraints.
Additionally, the investigated slightly extended version of the MPC from Section \ref{sec: MPC}, i.e. MPC 2, showed a predictive behavior and beneficial performance using piece-wise constant stage cost, terminal cost and terminal region.  
This allowed us to emphasize the ability to use information about future heat demand and temperature requirements in the context of an MPC.

\section{Conclusion} \label{sec: conclusion}
Given the significance of developing operating strategies for RES-based DHGs, we have identified MPC as a promising strategy that effectively amalgamates the benefits of existing optimization-based EMSs with the strengths of established control approaches for DHGs.
To this end, we derived an ODE-based model describing the thermo-hydraulics of DHGs containing TESs with variable storage mass ratio as well as Kirchhoff's laws for hydraulic networks.
We proved the stabilizability of the DHG model, which is a necessary system property to apply MPC with terminal ingredients.
Via a case study, we demonstrated asymptotic stabilization of different steady state set points of an exemplary DHG. 

In future work, we plan to investigate the impact of incorporating a piece-wise constant stage cost function into the MPC scheme and examine the scalability of our approach.

\section*{Acknowledgment}
This research received funding from the German Federal Government, the Federal Ministry of Education and Research, and the State of Brandenburg within the framework of the joint project EIZ: Energy Innovation Center (project numbers 85056897 and 03SF0693A).

\bibliographystyle{IEEEtran}
\bibliography{main}

\end{document}